\documentclass{iopart}
\usepackage{iopams}
\usepackage{graphicx,cite}
\begin{document}
\title[]{Universality of energy conversion efficiency for optimal tight-coupling heat engines and refrigerators}
\author{Shiqi Sheng and Z C Tu\footnote{corresponding author. E-mail: tuzc@bnu.edu.cn}}

\address{Department of Physics, Beijing Normal University, Beijing 100875, China}

\begin{abstract}
A unified $\chi$-criterion for heat devices (including heat engines and refrigerators) which is defined as the product of the energy conversion efficiency and the heat absorbed per unit time by the working substance [de Tom\'{a}s~\emph{et al} 2012 \textit{Phys. Rev. E} \textbf{85} 010104(R)] is optimized for tight-coupling heat engines and refrigerators operating between two heat baths at temperatures $T_c$ and $T_h~(>T_c)$. By taking a new convention on the thermodynamic flux related to the heat transfer between two baths, we find that for a refrigerator tightly and symmetrically coupled with two heat baths, the coefficient of performance (i.e., the energy conversion efficiency of refrigerators) at maximum $\chi$ asymptotically approaches to
$\sqrt{\varepsilon_C}$ when the relative temperature difference between two heat baths $\varepsilon_C^{-1}\equiv (T_h-T_c)/T_c$ is sufficiently small. Correspondingly, the efficiency at maximum $\chi$ (equivalent to maximum power) for a heat engine tightly and symmetrically coupled with two heat baths is proved to be $\eta_C/2+\eta_C^2/8$ up to the second order term of $\eta_C\equiv (T_h-T_c)/T_h$, which reverts to the universal efficiency at maximum power for tight-coupling heat engines operating between two heat baths at small temperature difference in the presence of left-right symmetry [Esposito \emph{et al} 2009 \textit{Phys. Rev. Lett.} \textbf{102} 130602].
\end{abstract}


%\maketitle

\section{Introduction}

One of key topics in finite-time thermodynamics is the efficiency at maximum power for heat engines, which has been widely investigated by many researchers~\cite{Yvon55,Novikov,Chambadal,Curzon1975,Andresen1977,Chen1989,ChenJC94,Bejan96,ChenL99,vdbrk2005,dcisbj2007,Schmiedl2008,Tu2008,Esposito2009,Apertet12,Seifert12rev,Izumida2012,Esposito2010,Esposito2009a,GaveauPRL10,WangTu2011,wangtu2012,WangHe,wangheinter}.
The most elegant result on this topic is $\eta_{CA}\equiv 1-\sqrt{1-\eta_C}$, the efficiency at maximum power for endoreversible heat engines~\cite{Yvon55,Novikov,Chambadal,Curzon1975}, where $\eta_C$ represents the Carnot efficiency. Recently, the efficiencies at maximum power for stochastic heat engines~\cite{Schmiedl2008}, the Feynman ratchet~\cite{Tu2008}, and quantum dot engines\cite{Esposito2009a} were respectively achieved, and their expressions look quite different from $\eta_{CA}$. It is this difference that attracts researchers' attention to the following two issues: One is the universal efficiency at maximum power for tight-coupling heat engines operating between two baths at small temperature difference \cite{vdbrk2005,dcisbj2007,Schmiedl2008,Tu2008,Esposito2009a,Esposito2009,Apertet12,Seifert12rev}; the other is the global bounds of efficiency at maximum power for heat engines operating between two baths with arbitrary temperature difference~\cite{Esposito2010,GaveauPRL10,WangTu2011,wangtu2012,Izumida2012,WangHe,wangheinter,wangtu13ctp}.
In particular, Van den Broeck~\cite{vdbrk2005} found that the universal efficiency at maximum power for tight-coupling heat engines is equal to $\eta_C/2$ up to the first order term of relative temperature difference between two baths. The universality up to the second order term of relative temperature difference was first observed in Ref.~\cite{Schmiedl2008}, then proposed as a conjecture in Ref.~\cite{Tu2008}, and finally proved by Esposito \emph{et al.} for tight-coupling heat engines in the presence of left-right symmetry~\cite{Esposito2009}. In addition, Esposito \emph{et al.} also found the efficiency at maximum power of heat engines to be bounded between $\eta_C/2$ and $\eta_C/(2-\eta_C)$ under low-dissipation conditions~\cite{Esposito2010}. Interestingly, these two bounds are also shared by the linear-irreversible heat engines \cite{WangTu2011}. The accessibility of the bounds for different kinds of heat engines were also investigated in Refs.~\cite{wangtu2012,Izumida2012,WangHe,wangheinter,wangtu13ctp}.

On the other hand, the maximum power point was proved to be unavailable for refrigerators~\cite{Andresen1977}, which leads to discussions~\cite{YanChen1990,Jincan1998,Jincanjpa09, Jizhouhe02,dcisbj2006,TRHWT2013,Mahler2010,RocoPRE12,Velasco1997,Lingenchen1995,ChenDingsun11,WLTHRpre12,Izumida13} on the proper optimization criterion for refrigerators and the corresponding coefficient of performance (COP). Yan and Chen~\cite{YanChen1990} suggested to take $\varepsilon \dot{Q}_c$ as the target function, where $\varepsilon$ and $\dot{Q}_c$ are the COP of refrigerators and the heat absorbed by the working substance from the cold bath per unit time, respectively. They also optimized this target function and found the corresponding COP to be $\varepsilon_{CY}\equiv\sqrt{\varepsilon_C+1}-1$ for endoreversible refrigerators~\cite{YanChen1990}, where $\varepsilon_C$ is the Carnot COP for reversible refrigerators. Recently, de Tom\'{a}s \emph{et al.} introduced a
unified target function for heat devices including heat engines and refrigerators. This target function is now called $\chi$-criterion which is defined as the product of the energy conversion efficiency and the heat absorbed per unit time by the working substance~\cite{RocoPRE12}. $\chi$-criterion reverts to the power output for heat engines
since the energy conversion efficiency for heat engines is exactly the thermal efficiency which is defined as the ratio of the power output to the heat absorbed per unit time by the working substance (from the hot bath). It also degenerates into the target function proposed by Yan and Chen for refrigerators
since the energy conversion efficiency is exactly the COP of refrigerators and the the working substance of refrigerators absorbs heat $\dot{Q}_c$ from the cold bath per unit time. Additionally, the COP at maximum $\chi$ was also found to be $\varepsilon_{CY}\equiv\sqrt{1+\varepsilon_C}-1$ for symmetric low-dissipation refrigerators~\cite{RocoPRE12}. Based on the work by de Tom\'{a}s \emph{et al.}, one of the present authors and his coworkers derived that the COP at maximum $\chi$ was bounded between $0$ and $(\sqrt{9+8\varepsilon_C}-3)/2$ for low-dissipation refrigerators~\cite{WLTHRpre12}. The observed COP's for real refrigerators are also located in the region between these bounds, which is in good agreement with their theoretical estimation. These bounds were also confirmed by Izumida \emph{et al.} with a minimally nonlinear irreversible model for refrigerators~\cite{Izumida13}.

The above researches on refrigerators support that $\chi$-criterion is an appropriate figure of merit for refrigerators. The results obtained from maximizing $\chi$ for refrigerators have counterparts in those derived from maximizing the power output (equivalent to the $\chi$-criterion) for heat engines. To begin with, the COP at maximum $\chi$-criterion ($\varepsilon_{CY}\equiv\sqrt{1+\varepsilon_C}-1$) for endoreversible refrigerators~\cite{YanChen1990} corresponds to the efficiency at maximum power ($\eta_{CA}\equiv 1-\sqrt{1-\eta_C}$) for endoreversible heat engines~\cite{Yvon55,Novikov,Chambadal,Curzon1975}. Next, the bounds [$0$ and $(\sqrt{9+8\varepsilon_C}-3)/2$] of COP at maximum $\chi$ for low-dissipation refrigerators~\cite{WLTHRpre12} correspond to the bounds [$\eta_C/2$ and $\eta_C/(2-\eta_C)$] of efficiency at maximum power for low-dissipation heat engines~\cite{Esposito2010}. In particular, the COP at maximum $\chi$-criterion ($\varepsilon_{CY}\equiv\sqrt{1+\varepsilon_C}-1$) for symmetric low-dissipation refrigerators also corresponds to the efficiency at maximum power ($\eta_{CA}\equiv 1-\sqrt{1-\eta_C}$) for symmetric low-dissipation heat engines~\cite{RocoPRE12}.
However, there is still a lack of the counterpart of universal efficiency at maximum power for tight-coupling heat engines in the COP at maximum $\chi$-criterion for tight-coupling refrigerators when they operate between two baths at small temperature difference. Our main goal in this work is to complement this theoretical imperfection. Based on the expression of COP at maximum $\chi$ for endoreversible refrigerators and symmetric low-dissipation refrigerators, we conjecture that $\sqrt{\varepsilon_C}$ might be the universal COP at maximum $\chi$ for tight-coupling refrigerators operating between two baths at small temperature difference. This conjecture is proved to be true under certain symmetric conditions within the framework of linear irreversible thermodynamics by adopting a new convention (eq.~\ref{eq-newconv}) on the thermodynamic flux related to the heat transfer between two baths. The universal behavior of efficiency at maximum power for tight-coupling heat engines is also recalculated with the consideration of this new convention.

\section{Generic models and new convention on thermodynamic flux}

\begin{figure}[htp!]\begin{center}
\includegraphics[width=8cm]{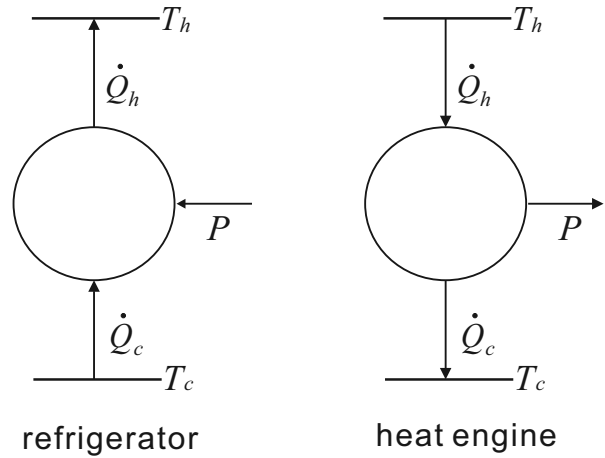}\caption{Generic setups for a tight-coupling refrigerator (left) and a tight-coupling heat engine (right).\label{fig-generic}}\end{center}
\end{figure}

As it was done by Van den Broeck~\cite{vdbrk2005} for heat engines, here we consider a generic setup for a tight-coupling refrigerator shown in Fig.~\ref{fig-generic}. An external force $F$ is applied on the system and inputs a power $P=F\dot{x}$ into the system, where $x$ is the thermodynamically conjugate variable of $F$. The dot represents the derivative with respect to time. The corresponding thermodynamic force is $X_{1}={F}/{T}$, where $T$ is the
temperature of the system which can be well defined due to the assumption of local equilibrium. The thermodynamic flux conjugated to $X_{1}$ is $J_1=\dot{x}$. Then the power input can be expressed as $P=J_{1}X_{1}T$ in terms of the thermodynamic flux and force.
Assume that the system is in contact with a cold bath at temperature $T_c\equiv T-s_c\Delta T$ and a hot bath at temperature $T_h\equiv T+s_h\Delta T$ with $\Delta T\ll T$. The positive parameters $s_c$ and $s_h$ should satisfy $s_c+s_h=1$ due to the constraint $\Delta T=T_h-T_c$. Their specific values depend on the coupling strengths between the model system and the cold or hot baths.

In steady state, since a certain amount of heat $\dot{Q}_h$ is released to the hot bath and heat $\dot{Q}_c$ is simultaneously absorbed from the cold bath by the refrigerator per unit time, the total entropy production rate can be expressed as $\sigma=\dot{Q}_h/T_h-\dot{Q}_c/T_c$, which can be further expressed as \begin{equation}\sigma=(\dot{Q}_c+P)/T_h-\dot{Q}_c/T_c={P}/{T_h}+\dot{Q}_c(1/T_h-1/T_c)\label{eq-e11}\end{equation}
or
\begin{equation}\sigma=\dot{Q}_h/{T_h}-(\dot{Q}_h-P)/{T_c}={P}/{T_c}+\dot{Q}_h(1/T_h-1/T_c)\label{eq-e12}\end{equation} due to the conservation of energy ($\dot{Q}_{c}+P=\dot{Q}_{h}$). Considering $T_h\equiv T+s_h\Delta T$ and $T_c\equiv T-s_c\Delta T$, we can transform (\ref{eq-e11}) and (\ref{eq-e12}) into
\begin{equation}\sigma =(P/T)[1-s_h\Delta T/T+ \mathcal{O}(\Delta T/T)^2]+\dot{Q}_c(1/T_h-1/T_c)\label{eq-e13}\end{equation}
and
\begin{equation}\sigma =(P/T)[1+s_c\Delta T/T+ \mathcal{O}(\Delta T/T)^2]+\dot{Q}_h(1/T_h-1/T_c),\label{eq-e14}\end{equation}
respectively, where $\mathcal{O}(\Delta T/T)^2$ represents the term in the same order of $(\Delta T/T)^2$.
Multiplying (\ref{eq-e13}) by $s_c$ and (\ref{eq-e14}) by $s_h$, respectively, then adding them up together, we can further derive
\begin{equation}\sigma=J_1X_1[1+ \mathcal{O}(\Delta T/T)^2]+(s_c\dot{Q}_c+s_h\dot{Q}_h)(1/T_h-1/T_c) \end{equation}
with the consideration of $s_c+s_h=1$ and $P=J_1X_1T$ for refrigerators.
The above equation enlightens us to take
\begin{equation}J_2\equiv s_c\dot{Q}_c+s_h\dot{Q}_h \label{eq-newconv}\end{equation}
and
\begin{equation}
X_2\equiv 1/T_h-1/T_c\label{eq-force2ref}\end{equation}
respectively as the thermodynamic flux and force related to the heat transfer between two heat baths.
The truncation error with the consideration of new convention (\ref{eq-newconv}) is in the order of $(\Delta T/T)^2$ when the entropy production rate is expressed as $\sigma =J_1X_1 +J_2 X_2$. The other conventions such as $J_2\equiv \dot{Q}_c$ or $J_2\equiv \dot{Q}_h$ result in the lower accuracy with the truncation error in the order of $\Delta T/T$ when the entropy production rate is expressed as $\sigma =J_1X_1+J_2 X_2$.

Now we turn to the generic setup for a tight-coupling heat engine as shown in Fig~\ref{fig-generic}. As was done by Van den Broeck~\cite{vdbrk2005}, we assume that the system outputs power $P=-F\dot{x}$ against an external force $F$, where $x$ is the thermodynamically conjugate variable of $F$. The corresponding thermodynamic force is $X_{1}={F}/{T}$, where $T$ is the
temperature of the system. The thermodynamic flux conjugated to $X_{1}$ is $J_1=\dot{x}$. Then the power output can be expressed as $P=-J_{1}X_{1}T$ in terms of the thermodynamic flux and force.
Assume that the system is in contact with a cold bath at temperature $T_c\equiv T-s_c\Delta T$ and a hot bath at temperature $T_h\equiv T+s_h\Delta T$ with $\Delta T\ll T$.

Similar to the analysis of refrigerators, since a certain amount of heat $\dot{Q}_h$ is absorbed from the hot bath and heat $\dot{Q}_c$ is simultaneously released to the cold bath by the heat engine per unit time, the total entropy production rate $\sigma={\dot{Q}_c}/{T_c}-{\dot{Q}_h}/{T_h}$ can be expressed as
\begin{equation}\sigma={\dot{Q}_c}/{T_c}-(\dot{Q}_c+P)/{T_h}=-{P}/{T_h}+\dot{Q}_c({1}/{T_c}-{1}/{T_h}),\label{eq-entgen5c}\end{equation}
or
\begin{equation}\sigma=(\dot{Q}_h-P)/{T_c}-{\dot{Q}_h}/{T_h}=-{P}/{T_c}+\dot{Q}_h({1}/{T_c}-{1}/{T_h}).\label{eq-entgen5d}\end{equation}
If considering $T_h\equiv T+s_h\Delta T$ and $T_c\equiv T-s_c\Delta T$, we can transform (\ref{eq-entgen5c}) and (\ref{eq-entgen5d}) into
\begin{equation}\sigma=-({P}/{T})[1-{s_h\Delta T}/{T}+\mathcal{O}({\Delta T}/{T})^2]+\dot{Q}_c ({1}/{T_c}-{1}/{T_h}),\label{eq-entgen5a}\end{equation}
and
\begin{equation}\sigma=-({P}/{T})[1+{s_c\Delta T}/{T}+\mathcal{O}({\Delta T}/{T})^2]+\dot{Q}_h({1}/{T_c}-{1}/{T_h}).\label{eq-entgen5b}\end{equation}
From the above two equations, we obtain
\begin{equation}\sigma=J_1X_1[1+\mathcal{O}({\Delta T}/{T})^2]+(s_c\dot{Q}_c+s_h\dot{Q}_h)({1}/{T_c}-{1}/{T_h})\label{eq-entgen5}\end{equation}
with the consideration of $s_c+s_h=1$ and $P=-J_1X_1T$ for heat engines.
The above equation enlightens us to take the thermodynamic flux $J_2\equiv s_c\dot{Q}_c+s_h\dot{Q}_h$ and the thermodynamic force $X_2\equiv{1}/{T_c}-{1}/{T_h}$ such that the entropy production rate can be expressed as $\sigma=J_1 X_1 +J_2 X_2$ with a truncation error in the order of $(\Delta T/T)^2$. We note that the expression of thermodynamic force $X_2$ for heat engines opposites to the thermodynamic force (\ref{eq-force2ref}) for refrigerators while the thermodynamic flux has the same form as (\ref{eq-newconv}).

\section{Universality of COP at maximum $\chi$ for tight-coupling refrigerators}

It is easy to see that $\varepsilon_{CY}\equiv\sqrt{\varepsilon_{C}+1}-1 \simeq \sqrt{\varepsilon_{C}}$ when $\varepsilon_{C}\rightarrow\infty$. That is, endoreversible refrigerators and symmetric low-dissipation refrigerators share the same asymptotic behavior at small temperature difference $\varepsilon_{C}^{-1}=(T_h-T_c)/T_c\ll 1$. Therefore, we may conjecture that a universal COP at maximum $\chi$-criterion,
\begin{equation}
\varepsilon_{\ast}\simeq \sqrt{\varepsilon_{C}},\label{eq-copsmalt}\end{equation} might exist for
tight-coupling refrigerators working between two baths at small temperature difference since both kinds of refrigerators can be regarded as tight-coupling refrigerators~\cite{tightproof}.

Now we will adopt the high-precision convention (\ref{eq-newconv}) to deal with tight-coupling refrigerators.
According to linear irreversible thermodynamics, we write the linear
relationship between the thermodynamic fluxes and forces:
\begin{equation}J_{1}=L_{11}X_{1}+L_{12}X_{2},~  J_{2}=L_{21}X_{1}+L_{22}X_{2},\label{eq-lirrevt}\end{equation}
where the Onsager coefficients satisfy $L_{11}\geq 0$, $L_{22}\geq 0$, $L_{11}L_{22}-L_{12}L_{21}\geq 0$ and $L_{12}=L_{21}$.
It is necessary to point out that the expression of $J_1$ implies that $P=TJ_1X_1$ is a quadratic form. Thus $\dot{Q}_h$ and $\dot{Q}_c$ may contain quadratic terms due to the constraint $\dot{Q}_h-\dot{Q}_{c}=P=TJ_1X_1$. It is unsuitable to merely take either $\dot{Q}_h$ or $\dot{Q}_c$ as $J_2$ because $J_2$ is a linear form. Our new convention (\ref{eq-newconv}) ensure $J_2$ to be linear through a special combination between $\dot{Q}_h$ and $\dot{Q}_c$, which is reflected in the truncation error in the order of $(\Delta T/T)^2$ when the entropy production rate is expressed as  $\sigma=J_1 X_1 +J_2 X_2$.

Furthermore, the tight-coupling condition $L_{12}^{2}=L_{21}^{2}=L_{11}L_{22}$ leads to
\begin{equation}J_2/J_1=L_{12}/L_{11},\label{eq-tccond}\end{equation}
which indicates that the mechanical flux is proportional to the heat flux under the tight-coupling condition~\cite{vdbrk2005}.

With the consideration of (\ref{eq-newconv}) and $\dot{Q}_h-\dot{Q}_{c}=P$, the heat absorbed from the cold bath can be expressed as $\dot{Q}_{c}=J_2-s_h P$. Thus, the COP can be expressed as $\varepsilon\equiv\dot{Q_{c}}/{P}={L_{21}}/{TL_{11}X_{1}}-s_h$ due to $P=J_1X_1T$, from which we have $X_{1}={L_{12}}/TL_{11}{(\varepsilon+s_h)}$.
Substituting this equation into the definition of $\chi$-criterion, we obtain
\begin{equation}\chi\equiv\varepsilon\dot{Q}_{c}=L_{22}\varepsilon^{2}[1+TX_{2}(\varepsilon+s_h)]/T(\varepsilon+s_h)^2.\end{equation}

Maximizing $\chi$ with respect to $\varepsilon$ (equivalently with respect to $X_1$), we find that the COP at maximum $\chi$-criterion satisfies $\varepsilon_\ast^{2}+{3s_h\varepsilon_\ast}+2s_h^2+2s_h/TX_{2}=0$,
which gives $\varepsilon_\ast=\sqrt{{s_h^2}/{4}-2s_h/TX_2}-{3s_h}/{2}$.
Considering $X_{2}\equiv 1/T_h-1/T_c$ and $\varepsilon_C\equiv T_c/(T_h-T_c)$,
we achieve
\begin{equation}\varepsilon_\ast=\sqrt{{s_h^2}/{4}+{2s_h\varepsilon_C(\varepsilon_C+1)}/{(\varepsilon_C+s_c)}}-{3s_h}/{2}.\label{eq-coprefig23}\end{equation}

If the model system is symmetrically coupled with both heat baths such that $s_h=s_c=1/2$, we derive $\varepsilon_\ast\simeq\sqrt{\varepsilon_C}$ when $\varepsilon_C\rightarrow \infty$ from (\ref{eq-coprefig23}). Therefore, we have proved the conjecture on the universal COP at maximum $\chi$ for tight-coupling refrigerators symmetrically interacting with two heat baths at small temperature difference. In addition, in the extremely asymmetric cases of $s_h=0$ (i.e., $s_c=1$) and $s_h=1$ (i.e., $s_c=0$), equation~(\ref{eq-coprefig23}) leads to $\varepsilon_{0}\equiv0$ and $\varepsilon_{1}\equiv(\sqrt{9+8\varepsilon_C}-3)/2$, respectively, which surprisingly equate the global lower and upper bounds [$\varepsilon_{-}\equiv0$ and $\varepsilon_{+}\equiv(\sqrt{9+8\varepsilon_C}-3)/2$] of COP at maximum $\chi$ for low-dissipation refrigerators which are also found to be reached in the case of extremely asymmetric dissipations~\cite{WLTHRpre12}. Indeed, equation~(\ref{eq-coprefig23}) implies that $\varepsilon_\ast$ is strictly bounded between $\varepsilon_{0}\equiv0$ and $\varepsilon_{1}\equiv(\sqrt{9+8\varepsilon_C}-3)/2$ if the temperature difference is not too large.

\section{Universality of efficiency at maximum $\chi$ for tight-coupling heat engines}

In this section, we adopt convention (\ref{eq-newconv}) to recalculate the efficiency at maximum maximum $\chi$ (i.e., maximum power) for tight-coupling heat engines.

The heat absorbed from the hot bath can be expressed as $\dot{Q}_h=J_2+s_cP$ with the consideration of $\dot{Q}_h=\dot{Q}_c+P$. So the efficiency can be expressed as
\begin{eqnarray}\eta\equiv \frac{P}{\dot{Q}_h}=\frac{P}{J_2+s_cP}=\frac{1}{s_c-L_{12}/L_{11}TX_1}\label{eq-effc}\end{eqnarray}
with the consideration of tight-coupling condition (\ref{eq-tccond}) and $P=-J_1X_1T$ for heat engines.

On the other hand, the $\chi$-criterion degenerates into the power output for heat engines~\cite{RocoPRE12}. Within the
linear irreversible thermodynamics, the power output can be expressed as
\begin{eqnarray}P=-TJ_1X_1=-T(L_{11}X_1+L_{12}X_2)X_1\end{eqnarray}
with the consideration of linear relationship (\ref{eq-lirrevt}).
Maximizing $P$ with respect to $X_1$, we obtain the optimized $X_1=-L_{12}X_2/2L_{11}$. Substituting it into (\ref{eq-effc}), we obtain the efficiency at maximum power
$\eta_\ast=TX_2/(2+s_cTX_2)$. By considering $X_2\equiv {1}/{T_c}-{1}/{T_h}$ and
$\eta_C\equiv(T_h-T_c)/T_h$, we can derive the efficiency at maximum power:
\begin{eqnarray}\eta_\ast=\frac{\eta_C(1-s_h\eta_C)}{2(1-\eta_C)+s_c\eta_C(1-s_h\eta_C)}=\frac{\eta_C}{2}+\frac{s_c\eta_C^2}{4}+\mathcal{O}(\eta^3_C).\label{eq-effcc3}\end{eqnarray}
Thus $\eta_\ast$ is equal to $\eta_C/2$ up to the first order term in the limit of small relative temperature difference as it was obtained in Ref.~\cite{vdbrk2005}.
Interestingly, when the model system is symmetrically coupled with both heat baths such that $s_c=s_h=1/2$, we arrive at
\begin{equation}\label{eq-univeta}
\eta_\ast\simeq \eta_C/2+\eta_C^2/8,\end{equation}
which equates the universal efficiency at maximum power up to the second order term of $\eta_C$ for tight-coupling heat engines in the presence of left-right symmetry \cite{Esposito2009}.

In addition, in the extremely asymmetric cases of $s_c=0$ (i.e., $s_h=1$) and $s_c=1$ (i.e., $s_h=0$), equation (\ref{eq-effcc3}) leads to $\eta_{0}\equiv\eta_C/2$ and $\eta_{1}\equiv \eta_C/(2-\eta_C)$, respectively, which surprisingly equate the global lower and upper bounds [$\eta_{-}\equiv\eta_C/2$ and $\eta_{+}\equiv \eta_C/(2-\eta_C)$] of efficiency at maximum power for low-dissipation heat engines which are also found to be reached in the case of extremely asymmetric dissipations~\cite{Esposito2010}. Indeed, equation~(\ref{eq-effcc3}) implies that $\eta_\ast$ is strictly bounded between $\eta_{0}\equiv\eta_C/2$ and $\eta_{1}\equiv \eta_C/(2-\eta_C)$ if the temperature difference is not too large.

\section{Conclusion and discussion}

In the above discussions, we have obtained the universal COP (\ref{eq-copsmalt}) at maximum $\chi$ for refrigerators tightly and symmetrically coupled with two baths at small temperature difference with the consideration of new convention (\ref{eq-newconv}) as the thermodynamic flux related to the heat transfer between two baths. By adopting this convention, we have also derived the universal efficiency (\ref{eq-univeta}) at maximum power up to the second order term for tight-coupling heat engines symmetrically interacting with two heat baths at small temperature difference. In this sense, asymptotic behavior (\ref{eq-copsmalt}) for refrigerators tightly and symmetrically coupled with two baths at small temperature difference can be regarded as the counterpart of universal efficiency (\ref{eq-univeta}) at maximum power for tight-coupling heat engines operating between two baths at small temperature difference in the presence of left-right symmetry.

As a byproduct, we surprisingly find that the global bounds of efficiency at maximum power for low-dissipation heat engines obtained in Ref.~\cite{Esposito2010} and the global bounds of COP at maximum $\chi$ for low-dissipation refrigerators obtained in Ref.~\cite{WLTHRpre12} can be achieved when we adopt new convention (\ref{eq-newconv}) to deal with tight-coupling heat engines or refrigerators in the extremely asymmetric cases. It is necessary for us to investigate whether this fact takes place merely by coincidence or due to some underlying reasons in the future work.

\ack
The authors are grateful to the financial support of Nature Science Foundation of China (Grant NO. 11075015). They also thank Yang Wang for carefully proofreading the manuscript.


\begin{thebibliography}{99}
\bibitem{tightproof}The mapping from the endoreversible refrigerators and the symmetric low-dissipation refrigerators into tight-coupling refrigerators will be investigated in a full paper.
\end{thebibliography}
\end{document}